\documentclass{ws-ijmpd}

\begin{document}

\markboth{Lixin Xu, Hongya Liu and Chengwu Zhang}{Reconstruction
of $5D$ Cosmological Models From Equation of State of Dark Energy}

%
\catchline{}{}{}{}{}
%

\title{Reconstruction of $5D$ Cosmological Models From Equation of State of Dark Energy}

\author{Lixin Xu\footnote{lxxu@dl.cn},
Hongya Liu\footnote{Corresponding author: hyliu@dlut.edu.cn },
Chengwu Zhang}

\address{Department of Physics, Dalian University of
Technology, Dalian, 116024, P. R. China}

\maketitle

\begin{history}
\received{Day Month Year} \revised{Day Month Year}
\end{history}

\begin{abstract}
We consider a class of five-dimensional cosmological solutions
which contains two arbitrary function $\mu(t)$ and $\nu(t)$. We
found that the arbitrary function $\mu(t)$ contained in the
solutions can be rewritten in terms of the redshift $z$ as a new
arbitrary function $f(z)$. We further showed that this new
arbitrary function $f(z)$ could be solved out for four known
parameterized equations of state of dark energy. Then the $5D$
models can be reconstructed and the evolution of the density and
deceleration parameters of the universe can be determined.
\end{abstract}

\keywords{Kaluza-Klein theory; cosmology}


\section{Introduction}\label{I}

Recent observations of high redshift Type Ia supernovae reveal
that our universe is undergoing an accelerated expansion rather
than decelerated expansion \cite{RS,TKB,Riess}. Meanwhile, the
discovery of Cosmic Microwave Background (CMB) anisotropy on
degree scales together with the galaxy redshift surveys indicate
$\Omega _{total}\simeq 1$ \cite{BHS} and $\Omega _{m}\simeq \left.
1\right/ 3$. All these results strongly suggest that the universe
is permeated smoothly by 'dark energy', which violates the strong
energy condition with negative pressure and causes the expansion
rate of the universe accelerating. The dark energy and
accelerating universe have been discussed extensively from
different points of view \cite{Quintessence,Phantom,K-essence}. In
principle, a natural candidate for dark energy could be a small
cosmological constant. However, there exist serious theoretical
problems: fine tuning and coincidence problems. To overcome the
coincidence problem, some self-interact scaler fields $\phi$ with
an equation of state (EOS) $w_{\phi}=p_{\phi}\left/
\rho_{\phi}\right.$ were introduced dubbed quintessence, where
$w_{\phi}$ is time varying and negative. Generally, the potentials
of the scalar field should be determined from the underlying
physical theory, such as Supergravity, Superstring/M-theory etc..
However, from the phenomenal level, one can also design many kinds
of potentials to solve the concrete problems
\cite{Quintessence,sahni}. Once the potentials are given, EOS
$w_{\phi}$ of dark energy can be found. On the other hand, the
potential can also be reconstructed from a given EOS $w_{\phi}$
\cite{GOZ}. That is, the forms of the scalar potential can be
determined from observational data. Although there are many kinds
of models for dark energy, one still knew little about it's
properties. And, one also needs some mechanism to distinguish
these different models. Therefore, one may wish to use the model
independent method to study the universe without specifying a
particular model for dark energy. That is, we can use
observational data to parameterize the EOS of dark energy, and
then to study the evolution of the universe directly.

The idea that our world may have more than four dimensions is due
to Kaluza \cite{Kaluza}, who unified Einstein's theory of General
Relativity with Maxwell's theory of Electromagnetism in a $5D$
manifold. In 1926, Klein reconsidered Kaluza's idea and treated
the extra dimension as a compact small circle topologically
\cite{Klein}. Afterwards, the Kaluza-Klein idea has been studied
extensively from different points of view. Among them, a kind of
theory called Space-Time-Matter (STM) theory, is designed to
incorporate the geometry and matter by Wesson and his
collaborators (for review, please see \cite{Wesson} and references
therein). In STM theory, our $4D$ world is a hypersurface embedded
in a $5D$ Ricci flat ($R_{AB}=0$) manifold, and all the matter in
our $4D$ world are induced from the extra dimension. This theory
is supported by Campbell's theorem \cite{Compbell} which says that
any analytical solution of Einstein field equation of $N$
dimensions can be locally embedded in a Ricci-flat manifold of
$\left(N+1 \right)$ dimensions. Since the matter are induced from
the extra dimension, this theory is also called induced matter
theory.

Within the framework of STM theory, a cosmological solution is
presented in \cite{LiuW} in which it was shown that the universe
is characterized by having a big bounce instead of a big bang. It
was also shown that both the radiation and matter dominated
cosmological models could be recovered from the solution. Further
studies of this solution include the embedding to brane models
\cite{Seahra}, the isometry with $5D$ black holes \cite{SearhaW},
the big bounce singularity \cite{XuLiu}, and the dark energy
models \cite{5DE,WLX}. The purpose of this paper is to study the
acceleration of the $5D$ solution. The solution contains two
arbitrary functions $\mu(t)$ and $\nu(t)$. We will show in Section
\ref{II} that one of these two arbitrary functions, $\mu(t)$,
plays a similar role as the potential $V\left(\phi\right)$ in
quintessence or phantom dark energy models. This enable us to
study the evolution of the $5D$ universe in a model independent
way. We will reconstruct the evolution of the $5D$ universe by
using four known parameterized methods. Section \ref{III} is a
short discussion.

\section{Dark energy in a class of five-dimensional cosmological
model}\label{II}

The $5D$ cosmological solution was originally given by Liu and
Mashhoon in 1995 \cite{Liu}. Then, in 2001, Liu and Wesson
\cite{LiuW} restudied the solution and showed that it describes a
cosmological model with a big bounce as opposed to a big bang. The
$5D$ metric of this solution reads
\begin{equation}
dS^{2}=B^{2}dt^{2}-A^{2}\left(
\frac{dr^{2}}{1-kr^{2}}+r^{2}d\Omega ^{2}\right) -dy^{2}
\label{5-metric}
\end{equation}
where $d\Omega ^{2}\equiv \left( d\theta ^{2}+\sin ^{2}\theta
d\phi ^{2}\right) $ and
\begin{eqnarray}
A^{2} &=&\left( \mu ^{2}+k\right) y^{2}+2\nu y+\frac{\nu
^{2}+K}{\mu ^{2}+k},
\nonumber \\
B &=&\frac{1}{\mu }\frac{\partial A}{\partial t}\equiv
\frac{\dot{A}}{\mu }. \label{A-B}
\end{eqnarray}
Here $\mu =\mu (t)$ and $\nu =\nu (t)$ are two arbitrary functions
of $t$, $k$ is the $3D$ curvature index $\left(k=\pm 1,0\right)$,
and $K$ is a constant. This solution satisfies the 5D vacuum
equation $R_{AB}=0$. So, the three invariants are
\begin{eqnarray}
I_{1} &\equiv &R=0, I_{2}\equiv R^{AB}R_{AB}=0,
\nonumber  \\
I_{3} &=&R_{ABCD}R^{ABCD}=\frac{72K^{2}}{A^{8}}. \label{3-invar}
\end{eqnarray}
The invariant $I_{3}$ in Eq. (\ref{3-invar}) shows that $K$
determines the curvature of the 5D manifold.

Using the $4D$ part of the $5D$ metric (\ref{5-metric}) to
calculate the $4D$ Einstein tensor, one obtains
\begin{eqnarray}
^{(4)}G_{0}^{0} &=&\frac{3\left( \mu ^{2}+k\right) }{A^{2}},
\nonumber \\
^{(4)}G_{1}^{1} &=&^{(4)}G_{2}^{2}=^{(4)}G_{3}^{3}=\frac{2\mu \dot{\mu}}{A%
\dot{A}}+\frac{\mu ^{2}+k}{A^{2}}.  \label{einstein}
\end{eqnarray}
In Ref. \cite{WLX}, the induced matter was set to contain three
components: dark matter, radiation and $x$-matter. In this paper,
we assume, for simplicity, the induced matter to contain two
parts: cold dark matter (CDM) $\rho_{cd}$ and dark energy (DE)
$\rho_{de}$. So, we have
\begin{eqnarray}
\frac{3\left( \mu ^{2}+k\right) }{A^{2}} &=&\rho_{cd}+\rho_{de},  \nonumber \\
\frac{2\mu \dot{\mu}}{A\dot{A}}+\frac{\mu ^{2}+k}{A^{2}}
&=&-\left(p_{cd}+p_{de}\right), \label{FRW-Eq}
\end{eqnarray}
where
\begin{equation}
p_{cd}=0, \quad p_{de}=w_{de}\rho_{de}. \label{EOS-X}
\end{equation}
From Eqs.(\ref{FRW-Eq}) and (\ref{EOS-X}), one obtains the EOS of
the dark energy
\begin{equation}
w_{de}=\frac{p_{de}}{\rho_{de}}=-\frac{2\left. \mu
\dot{\mu}\right/ A \dot{A}+\left. \left( \mu ^{2}+k\right) \right/
A^{2}}{3\left. \left( \mu ^{2}+k\right) \right/
A^{2}-\rho_{cd0}A^{-3}},\label{wx}
\end{equation}
and the dimensionless density parameters
\begin{eqnarray}
\Omega _{cd} &=&\frac{\rho _{cd}}{\rho_{cd}+\rho_{de}}=\frac{\rho
_{cd0}}{3\left( \mu ^{2}+k\right) A},  \label{omega-cd} \\
\Omega_{de} &=&1-\Omega_{cd}. \label{omega-de}
\end{eqnarray}
where $\rho _{cd0}=\bar{\rho}_{cd0}A_{0}^{3}$ ($\bar{\rho}_{cd0}$
and $A_{0}$ denote the current density of CDM and scale factor at
present time, respectively (The subscript $0$ denotes value at
present time), and $\Omega_{cd}$ and $\Omega_{de}$ are
dimensionless density parameters of CDM and DE, respectively. The
Hubble parameter and deceleration parameter should be given as
\cite{LiuW}, \cite{WLX},
\begin{eqnarray}
H&\equiv&\frac{\dot{A}}{A B}=\frac{\mu}{A} \\
q \left(t, y\right)&\equiv&\left.
-A\frac{d^{2}A}{d\tau^{2}}\right/\left(\frac{dA}{d\tau}\right)^{2}
=-\frac{A \dot{\mu}}{\mu \dot{A}}, \label{df}
\end{eqnarray}
from which we see that $\dot{\mu}\left/\mu\right.>0$ represents an
accelerating universe, $\dot{\mu}\left/\mu\right.<0$ represents a
decelerating universe. So the function $\mu(t)$ plays a crucial
role in defining the properties of the universe at late time.
In this paper, we consider the spatially flat $k=0$ cosmological
model. From equations (\ref{wx})-(\ref{df}), it is easy to see
that these equations do not contain the second arbitrary $\nu(t)$
explicitly. So if we use the relation
\begin{eqnarray}
A_{0}\left/A \right.=1+z \label{A(z)}
\end{eqnarray}
and define $\mu_{0}^{2}\left/ \mu_{z}^{2}\right.=f\left(z\right)$
with $f(0)\equiv 1$, then these equations (\ref{wx})-(\ref{df})
can be expressed in terms of redshift $z$ as
\begin{eqnarray}
w_{de} &=&-\frac{1+\left(1+z\right)d\ln
f\left(z\right)\left/dz\right.}{3-3\Omega_{cd}}, \label{wx-2} \\
\Omega_{cd}&=&\Omega_{cd0}\left(1+z\right)f\left(z\right),\label{omega-cd-2} \\
\Omega_{de}&=&1-\Omega_{cd},\label{omega-de-2}\\
H^2&=&H_0^2(1+z)^2f(z)^{-1}, \label{friedmann}\\
q&=&\frac{1+3\Omega_{de}w_{de}}{2}=-\frac{\left(1+z\right)}{2}\frac{d\ln
f\left(z\right)}{dz}. \label{q}
\end{eqnarray}
Note that in the $5D$ bounce model the scale factor $A$ reaches a
nonzero minimum at $t=t_{b}$ where $t_{b}$ is the bouncing time.
From Eq. (\ref{A(z)}), this $t_{b}$ corresponds to a maximum
redshift $z_{b}$. Therefore, the relation (\ref{A(z)}) and all the
equations after it only valid in the range $z<z_{b}$ (i.e., after
the bounce).

Now let us consider equation (\ref{wx-2}) which is a first order
ordinary differential equation of the function $f(z)$ w.r.t.
redshift $z$. This equation could be integrated if the form of
$w_{de}(z)$ is given. It was shown in \cite{GOZ} that the scalar
potentials can be constructed from a given EOS of dark energy
$w_{\phi}$. Following this spirit, we can also reconstruct the
forms of function $f(z)$ from a given concrete form of
$w_{de}(z)$. And once the function $f(z)$ is constructed, the
evolution of the universe can be determined. At this point, we say
that the cosmological models are reconstructed.

Following Ref.\cite{GOZ}, we consider the following four cases to
reconstruct the forms of $f(z)$.

\begin{figure}
\begin{center}
\epsfbox{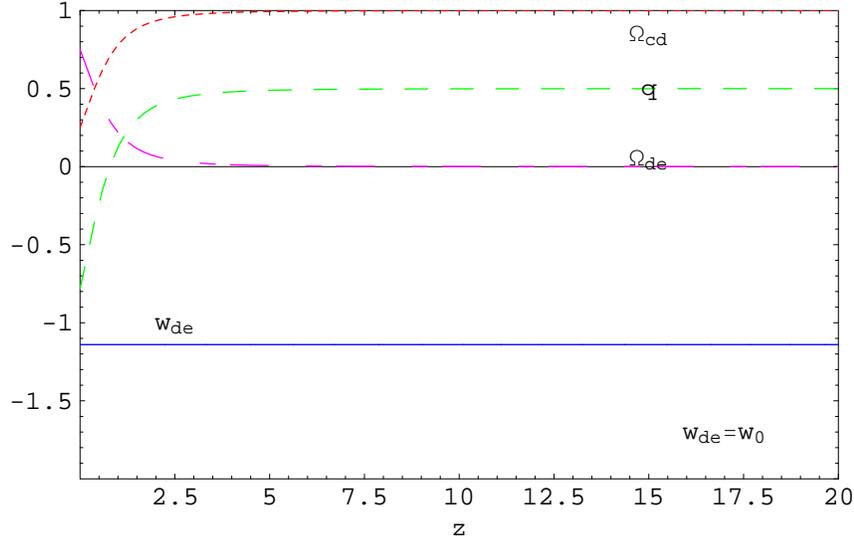}
\end{center}
\caption{{\bf Case I:} The evolution of the dimensionless density
parameters $\Omega_{cd}$, $\Omega_{de}$, and deceleration
parameter $q$, EOS of dark energy $w_{de}$ versus redshift $z$,
where $\Omega_{cd0}=0.25$, $\Omega_{de 0}=0.75$, and
$\omega_0=-1.14$.} \label{figz1}
\end{figure}

{\bf Case I:} $w_{de}=w_0$ (Ref. \cite{SHEM}) For this case,
$w_{de}$ is a constant and we find that Eq. (\ref{wx-2}) can be
integrated, giving
\begin{eqnarray}
f(z)=\frac{1}{(1+z)\left(\Omega_{cd0}+\Omega_{de0}(1+z)^{3
w_0}\right)}.\label{fz1}
\end{eqnarray}
Then using this $f(z)$ in (\ref{omega-cd-2}), (\ref{omega-de-2})
and (\ref{q}), we obtain $\Omega_{cd}$, $\Omega_{de}$ and
$\Omega_q$ expressed in terms of $z$ alone. The evolutions of the
dimensionless energy density parameters $\Omega_{cd}$ and
$\Omega_{de}$, EOS of dark energy $\omega_{de}$, and the
deceleration parameter $q$ are plotted in Fig. (\ref{figz1}).

\begin{figure}
\begin{center}
\epsfbox{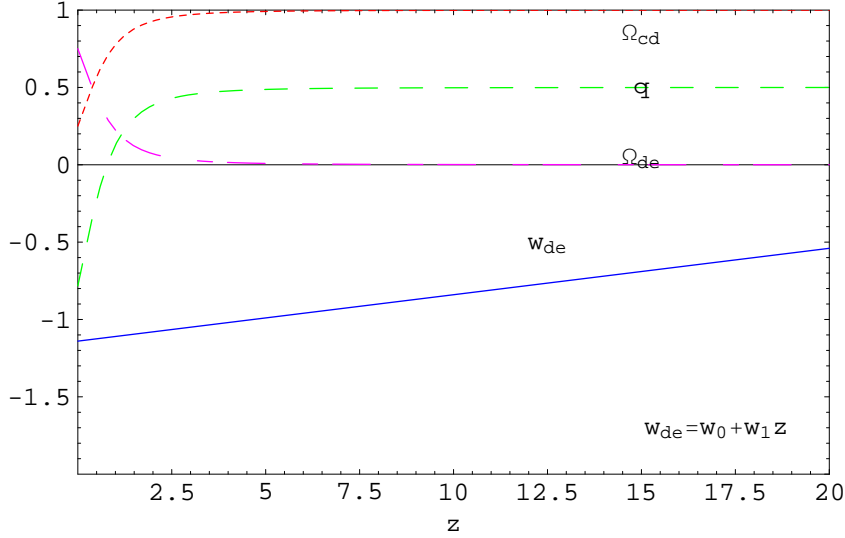}
\end{center}
\caption{{\bf Case II:} The evolution of the dimensionless density
parameters $\Omega_{cd}$, $\Omega_{de}$, and deceleration
parameter $q$, EOS of dark energy $w_{de}$ versus redshift $z$,
where $\Omega_{cd0}=0.25$, $\Omega_{de 0}=0.75$, $\omega_0=-1.14$
and $\omega_1=0.03$.} \label{figz2}
\end{figure}

{\bf Case II:} $w_{de}=w_0+w_1 z$ (Ref. \cite{ARCDH}) For this
case Eq. (\ref{wx-2}) can also be integrated, giving
\begin{equation}
f(z)=\frac{(1+z)^{3 w_1-1}}{\Omega_{cd0}(1+z)^{3
w_1}+\Omega_{de0}(1+z)^{3 w_0}\exp(3 w_1 z)}. \label{fz2}
\end{equation}
Using this $f(z)$ in (\ref{omega-cd-2}), (\ref{omega-de-2}) and
(\ref{q}), we obtain the expressions of the evolutions of the
dimensionless energy density parameters $\Omega_{cd}$ and
$\Omega_{de}$, EOS of dark energy $\omega_{de}$, and the
deceleration parameter $q$, and we plot them in Fig.
(\ref{figz2}).

\begin{figure}
\begin{center}
\epsfbox{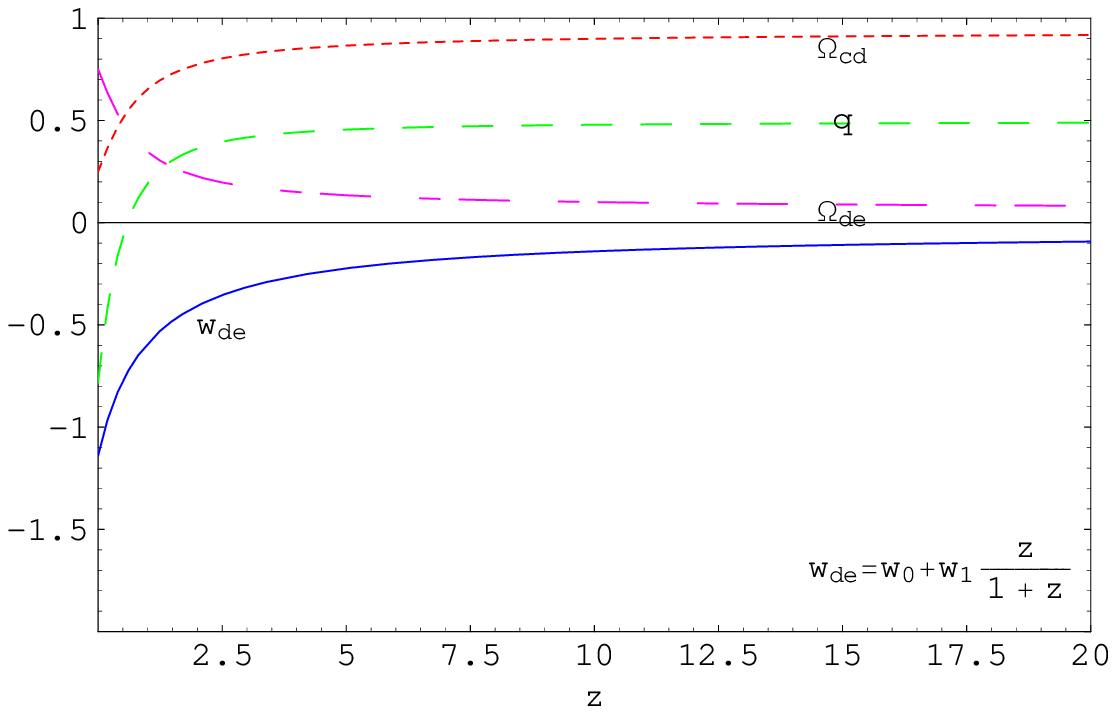}
\end{center}
\caption{{\bf Case III:} The evolution of the dimensionless
density parameters $\Omega_{cd}$, $\Omega_{de}$, and deceleration
parameter $q$, EOS of dark energy $w_{de}$ versus redshift $z$,
where $\Omega_{cd0}=0.25$, $\Omega_{de 0}=0.75$, $\omega_0=-1.14$
and $\omega_1=1.1$.} \label{figz3}
\end{figure}

{\bf Case III:} $w_{de}=w_0+w_1\frac{z}{1+z}$ (Ref. \cite{EVL,TP})
Similar as in case I and case II, for this case we have
\begin{equation}
f(z)=\frac{1}{(1+z)\left[\Omega_{cd0}+\Omega_{de0}(1+z)^{3w_0+3w_1}\exp(-\frac{3w_1
z}{1+z})\right]}.\label{fz3}
\end{equation}
We plot the evolutions of the dimensionless energy density
parameters $\Omega_{cd}$ and $\Omega_{de}$, EOS of dark energy
$\omega_{de}$ and deceleration parameter $q$ in Fig.
(\ref{figz3}).

\begin{figure}
\begin{center}
\epsfbox{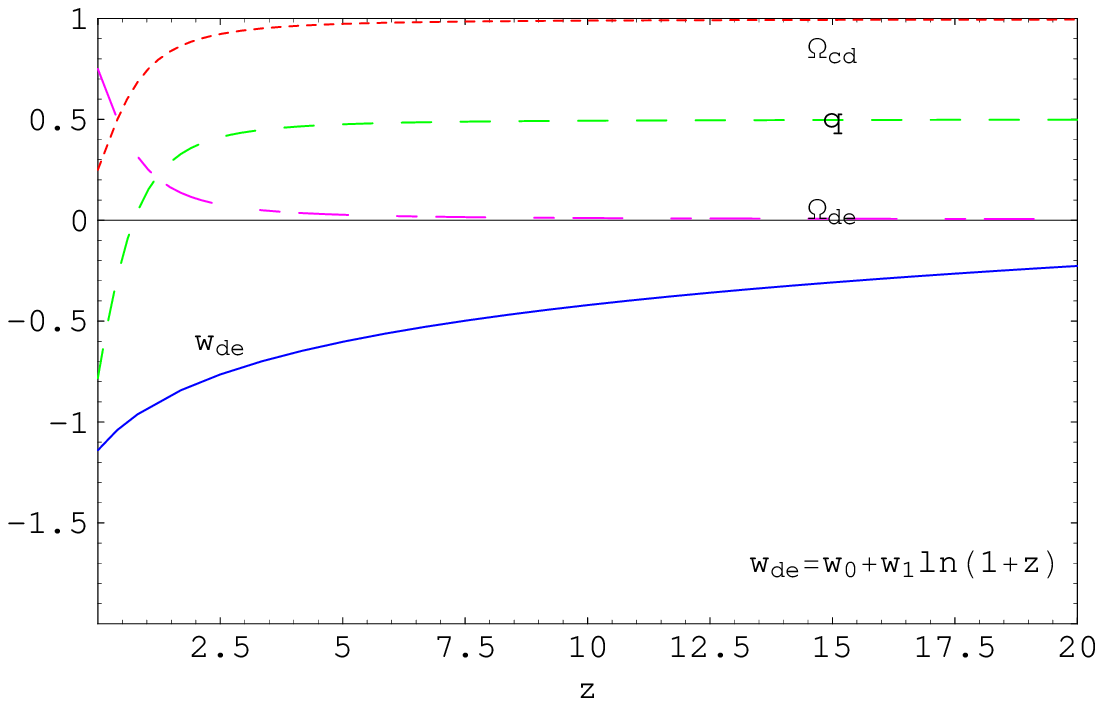}
\end{center}
\caption{{\bf Case IV:} The evolution of the dimensionless density
parameters $\Omega_{cd}$, $\Omega_{de}$, and deceleration
parameter $q$, EOS of dark energy $w_{de}$ versus redshift $z$,
where $\Omega_{cd0}=0.25$, $\Omega_{de 0}=0.75$, $\omega_0=-1.14$
and $\omega_1=0.3$.} \label{figz4}
\end{figure}

{\bf Case IV:} $w_{de}=w_0+w_1\ln(1+z)$ (Ref. \cite{BFGGE}) For
this case $f(z)$ is also integrable and we find
\begin{equation}
f(z)=\frac{1}{(1+z)\left[\Omega_{cd0}+\Omega_{de0}(1+z)^{3w_0}\exp(\frac{3w_1\ln(1+z)^2}{2})\right]}.\label{fz4}
\end{equation}
We plot the evolutions of the dimensionless energy density
parameters $\Omega_{cd}$ and $\Omega_{de}$, EOS of dark energy
$\omega_{de}$ and deceleration parameter $q$ in Fig.
(\ref{figz4}).

We see that in Case I the EOS of dark energy $w_{de}$ is assumed
to be a constant $w_0$ during the whole evolution of the universe.
In the other three cases, $w_{de}$ deviates from $w_0$ in
different ways as the redshift $z$ increases. This causes the
density and deceleration parameters $\Omega_{cd}$, $\Omega_{de}$
and $q$ deviate from those in Case I explicitly at higher
redshift. It is expected that these deviations may become very
large at very large redshift. In further studies we are going to
use more observational dada such as those from the SNe Ia data to
constrain the parameters in the expression of $w_{de}$. Here, in
this paper, we just want to use above four cases to illustrate how
to reconstruct function $f(z)$ from a given EOS of dark energy,
and the results show that our procedure works.

\section{Discussion}\label{III}

The $5D$ cosmological solution presented by Liu, Mashhoon and
Wesson in \cite{Liu} and\cite{LiuW} is rich in mathematics because
it contains two arbitrary functions $\mu(t)$ and $\nu(t)$.
However, this also brings us a problem: how to determine the two
arbitrary functions. In this paper we find that one of the two
functions, $\mu(t)$, plays a similar role as the potential
$V(\phi)$ in the quintessence and phantom dark energy models.
Meanwhile, another arbitrary function $\nu(t)$ seems do not affect
the densities and the EOS of dark energy in an explicit way. This
reminds us of a similar situation happened in the $4D$ general
relativity where one can study the cosmic evolution of dark energy
(as well as other densities) just from a given parameterized EOS
of dark energy without knowing it's explicit form and without
knowing the explicit form of the scale factor $a(t)$. Following
this kind of model independent methods we have used the relation
(\ref{A(z)}) and successfully derived the differential equation
(\ref{wx-2}) which governs the function $f(z)$. Furthermore, for
four known EOS of dark energy, we have successfully integrated Eq.
(\ref{wx-2}) and plotted the evolutions of the density and
deceleration parameters $\Omega_{cd}$, $\Omega_{de}$ and $q$. In
this sense, we say that we have reconstructed the $5D$ solution.
However, we should also say that this kind of reconstruction is
not complete. As we mentioned in Section \ref{II} that the
relation $A(z)=A_{0}(1+z)^{-1}$ in (\ref{A(z)}) does not cover the
whole $5D$ manifold of the bounce solution; it can only cover
"half" of it, i.e., that half after the bounce for $z<z_{b}$.
There is actually a one to two correspondence between $A$ and $t$
for the whole curve of $A$ in the bounce model; while the relation
$A(z)=A_{0}(1+z)^{-1}$ is just a one to one correspondence between
$A$ and $z$. On the other hand, even if in the $4D$ general
relativity, the scale factor $a(t)$ is not easy to be integrated
out analytically if the cosmic matter contains more than one
different components of matter. In our $5D$ case, the scale factor
$A(t,y)$ might be more complicated than $a(t)$. To study the
evolution of the whole bounce model, one may need look for other
method such as the numerical simulation which beyond the scope of
this paper.

\section{Acknowledgments}
This work was supported by NSF (10273004) and NBRP (2003CB716300)
of P. R. China.

\end{document}